\documentclass[prl,aps,showpacs,twocolumn,superscriptaddress]{revtex4}
\usepackage{graphicx}
\usepackage{epsfig}
\usepackage{amssymb}
\usepackage{amsmath}
\usepackage{amsfonts}
\usepackage{color}
\usepackage{bm}

\begin{document}

\title{Universality in the dynamics of second-order phase transitions}
\author{G. Nikoghosyan}
\affiliation{Institut f\"ur Theoretische Physik, Albert-Einstein Allee 11, Universit\"at
Ulm, 89069 Ulm, Germany}
\affiliation{Institute of Physical Research, 378410, Ashtarak-2, Armenia}
\author{R. Nigmatullin}
\affiliation{Department of Materials, University of Oxford, Oxford OX1 3PH, UK }

\author{M.B. Plenio}
\affiliation{Institut f\"ur Theoretische Physik, Albert-Einstein Allee 11, Universit\"at
Ulm, 89069 Ulm, Germany}

\begin{abstract}
When traversing a symmetry breaking second order phase transition at a finite rate, topological defects
form whose number dependence on the quench rate is given by simple power laws. We propose a general approach for the derivation
of such scaling laws that is based on the analytical transformation of the associated equations of motion
to a universal form rather than employing plausible physical arguments. We demonstrate the power of this
approach by deriving the scaling of the number of topological defects in both homogeneous and non-homogeneous
settings. The general nature and extensions of this approach are discussed.

\end{abstract}

\maketitle

{\em Introduction --} The study of the non-equilibrium dynamics of systems undergoing phase transitions is an
important problem of statistical physics \cite{doi:10.1080/00018739400101505,RevModPhys.49.435}. Of particular
interest in this context is the analysis of the dynamics of systems undergoing slow quenches through
a second-order phase transition from a symmetric to a symmetry broken phase \cite{Kibble2007}. When
the quench is performed at finite rates the symmetry is broken locally, and spatially separated regions
can select different states within the ground state manifold. The typical size of the correlated regions
and hence the density of defects exhibit a power-law dependence on the quench rate. One approach for the
prediction of the defect scaling, suggested in a theory which has become known as Kibble-Zurek
mechanism \cite{Kibble1980183,Zurek1985,Zurek1993}, employs physically reasonable arguments involving
the equilibrium concepts of divergence of correlation length, relaxation and freeze-out time near the
critical point. Indeed, for homogeneous and large systems approaching the thermodynamic limit, Kibble-Zurek
theory successfully predicts the scaling of defects with quench rate in terms of the critical exponents
of the phase transition. This intuitive derivation may indeed be transparent for such systems but is much
less so in more general cases involving for example those that exhibit spatial inhomogeneity
\cite{PhysRevLett.105.075701,1367-2630-12-11-115003}
which, in turn, poses the risk of incorrect conclusions to be drawn.
Indeed, while some experiments aimed at measuring defect scaling laws were realized in systems that are
well approximated as large and homogeneous systems, e.g. liquid crystals \cite{CHUANG15031991,Bowick18021994,PhysRevLett.83.5030}, liquid Helium \cite{Ruutu1996,Bauerle1996},
superconducting films \cite{PhysRevLett.91.197001,PhysRevLett.90.257001} and multiferroics \cite{PhysRevX.2.041022,PhysRevLett.108.167603}, the consequences of inhomogeneity and finite-size effects
cannot be ignored in recent experiments on Bose-Einstein condensates \cite{Sadler2006,Weiler2008,Lamporesi2013}
and ion crystals, \cite{Pyka2013,Ulm2013} which can exhibit non-standard scaling. Hence it is important to develop an
approach that does not appeal to physical intuition but exclusively relies on mathematical arguments in its
derivation of scaling laws. This will be the main goal and result of the present Letter.

{\em The basic idea --} Due to the application of a quench, that is an external change of the system parameters, the parameters in the
equations of motion adopt spatial and temporal dependencies whose rate of change is related to the quench
time $\tau_q$. The principal goal is the derivation of scaling laws for physical properties such as the
rate of defect formation in terms of $\tau_q$ {\it without} the explicit solution of the equations of motion
and {\it without} resorting to physical arguments concerning equilibrium concepts such as the divergence of
correlation length, relaxation and freeze-out time near the critical point. This goal is achieved by employing
as the principal idea the rescaling of the parameters of the equations of motion in such a manner that the
dependence on the quench rate is eliminated. The scaling of the spatial properties of the system is then
contained within this transformation. As this method uses purely the structure of the equations of motion,
it can be applied to a wide variety of systems.
For homogeneous systems in the thermodynamic limit our technique predicts scaling laws that are consistent
with the Kibble-Zurek theory in the mean-field regime. For the case of inhomogeneous and also finite size
systems we show that the precise scaling can be derived,
and it is not only a function of quench rate but also of the characteristic system size. We illustrate the
technique by applying it to power-law quenches in systems obeying the overdamped and underdamped time-dependent
Ginzburg-Landau equations but note that this approach is applicable well beyond Landau theory of second-order phase
transitions.

{\em Ginzburg-Landau Theory --} %
Let us start by introducing the free energy of the system in the vicinity of
the phase transition. For concreteness we assume that the state of the system is governed by
a complex order parameter $\phi\left( \mathbf{r},t\right)$, but the same arguments would apply for a real muticomponent order parameter. In Landau theory of phase transitions, the free energy of our system near the
critical point is given by
\begin{equation}
    \mathcal{F}_{tot}=\int d\mathbf{r}\left[ F_{grad}+F_{L}+F_{inh}\right] .
    \label{Free_energy}
\end{equation}
The first term on the rhs of (\ref{Free_energy}) is the gradient%
\begin{equation}
    F_{grad}=\frac{1}{2}\left\vert \nabla \phi\right\vert ^{2},
    \label{Free_energy2}
\end{equation}
the second term is the time-dependent Landau free energy
\begin{equation}
    F_{L}=\frac{1}{2} \varepsilon  \left( t\right) \left\vert \phi\right\vert ^{2}+%
    \frac{g}{4}\left\vert \phi\right\vert ^{4}  \label{Free_energy3}
\end{equation}%
and the last term represents the inhomogeneous external potential
\begin{equation}
    F_{inh}=\frac{1}{2}V\left( \mathbf{r},L\right) \left\vert \phi\right\vert ^{2}.
    \label{Free_energy4}
\end{equation}%
Here, $\varepsilon \left( t\right) $ is the time dependent critical parameter and
$L>0$ is the characteristic system size. For $\varepsilon + V\left( \mathbf{r},L\right) > 0$
the minimum of $F_{inh}+F_{L}$ corresponds to $\phi=0$, that is the field vanishes
and the system is in the symmetric phase. For $\varepsilon +V\left( \mathbf{r},L\right) <0$
there are many minima of $F_{inh}+F_{L}$ at $\left\vert \phi\left( \mathbf{r},t\right)
\right\vert ^{2}=-\frac{\varepsilon  +V\left( \mathbf{r},L\right) }{g}$ and
the system can 
adopt symmetry-broken states within this nontrivial ground-state manifold.
According to Landau theory, the phase transition takes place when the critical parameter
$\varepsilon \left(t\right) +V\left( \mathbf{r},L\right) $ changes its sign.

{\em The spatially homogeneous case --} In a first example we illustrate our method by considering
the equation of motion corresponding to phenomenological model A of \cite{RevModPhys.49.435}. The
evolution of the field is described by the
stochastic Ginzburg-Landau equation (SGLE)
\begin{equation}
    \frac{\partial \phi}{\partial t}=-\Gamma \frac{\delta \mathcal{F}_{tot}%
    }{\delta \phi }+\theta \left( \mathbf{r},t\right) \sqrt{T},
    \label{GL_eq}
\end{equation}
where $\Gamma $ is the relaxation, $T$ is the temperature of the system, and $\theta \left( \mathbf{r},t\right) $
is the uncorrelated white-noise variable with $\left\langle \theta \left( \mathbf{r},t\right) \theta^{*}
\left( \mathbf{r}^{\prime },t^{\prime }\right) \right\rangle = 2 \Gamma \delta \left( \mathbf{r}
-\mathbf{r}^{\prime }\right) \delta \left(t-t^{\prime }\right) $. The equation describing the evolution of the complex conjugate of the order parameter, $\phi^{*}(\textbf{r},t)$, is of the same form as eq. (\ref{GL_eq}).
The use of an uncorrelated white-noise
environment may be justified by taking the viewpoint that the physical environment couples to the microscopic
degrees of freedom of the system whose evolution is faster than that of the macroscopic order parameter
$\phi$. Hence, on the level of the macroscopic order parameter the environment is well approximated
by a Markovian environment. SGLE can be rigorously derived from the microscopic theory of the Bose-Einstein condensates \cite{Gardiner2003,Blakie2008}. It was used to model spontaneous defect formation in BEC \cite{Weiler2008}, giving good agreement with experimental observations.

First, we consider the dynamics of the system in a homogeneous
setting, i.e. $F_{inh}\equiv0$. By substituting  eqs. (\ref{Free_energy}), (\ref{Free_energy2})
and (\ref{Free_energy3}) into (\ref{GL_eq}) we obtain the following equation
of motion for the field
\begin{equation}
    \frac{\partial \phi}{\partial t}=-\Gamma \left[ \nabla ^{2}%
    \phi+\varepsilon \left( t\right) \phi+g\left\vert \phi%
    \right\vert ^{2}\phi\right] +\theta\left( \mathbf{r},t\right)
    \sqrt{T}.  \label{GP}
\end{equation}

In order to simplify the discussion we consider only polynomial dynamical quench
functions. It should be noted here, that other functional dependencies of quench
parameters can also be analyzed, but the analytical formulation of the result is
considerably more involved. Without loss of generality we assume that the phase
transition takes place at $t = 0$ and, furthermore, all quantities in the equations
of motion are assumed to be in dimensionless natural units. Therefore, the critical
parameter can be expressed as
\begin{equation}
    \varepsilon \left( t\right) =-\left\vert \dfrac{t}{\tau _{q}}\right\vert ^{n}%
    \mathrm{sign}\left( t\right) \label{eq:quench}
\end{equation}%
where $\tau_{q}$ defines the quench time scale. 

In general, the variation of the quench rate modifies the equation of motion (\ref{GP}) 
and thus affects the final state of the system. However, as\ we will show now, the 
dependence on $\tau _{q}$ can be eliminated and the equation of motion of the order 
parameter can be written in a universal form, if the parameters of equation (\ref{GP}) 
are rescaled. This is achieved by the consideration of the following linear transformations
\begin{align}
    \eta &=\alpha t,\text{ }\boldsymbol{\xi }=\beta \mathbf{r},\text{ }\text{ }\varphi=\frac{\phi}{\beta }
    \label{rescaling1}
\end{align}
and a suitable choice of $\alpha$ and $\beta$. Substituting rescalings (\ref{rescaling1}) 
in eq. (\ref{GP}) yields

\begin{align}
\frac{\partial \varphi}{\partial \eta }&=-\Gamma \left[
\frac{\beta^2}{\alpha}\nabla _{\xi }^{2}\varphi-\frac{1}{\alpha^{n+1}\tau_q^n}\mathrm{sign}\left( \eta \right) \left\vert
\eta \right\vert ^{n}\varphi +\right.\nonumber \\
 &  \left. +\frac{\beta^2}{\alpha}g\left\vert \varphi\right\vert ^{2}%
\varphi\right] +\sqrt{\frac{T\beta^{d-2}}{\alpha}}\theta\left( \boldsymbol{\xi },\eta \right),
\label{GP1_1}
\end{align}%
where $d$ is the dimensionality of the system. The dependence of the terms in $\left[ \;\right]$ in eq. (\ref{GP1_1}) on $\tau_q$ can be eliminated by choosing

\begin{align}
    \alpha &=\tau _{q}^{-\frac{n}{n+1}}, \text{ }\beta = \alpha^\frac{1}{2}=\tau _{q}^{-\frac{n}{2\left( n+1\right) }}.
    \label{rescaling2}
\end{align}
An additional rescaling of temperature, $\widetilde{T}=T\beta^{d-4}$, results in a fully $\tau_q$-invariant equation of motion
\begin{equation}
\frac{\partial \varphi}{\partial \eta }=-\Gamma \left[
\nabla _{\xi }^{2}\varphi-\mathrm{sign}\left( \eta \right) \left\vert
\eta \right\vert ^{n}\varphi+g\left\vert \varphi\right\vert ^{2}%
\varphi\right] +\sqrt{\widetilde{T}}\theta\left( \boldsymbol{\xi },\eta \right).  \label{GP2}
\end{equation}

Equation (\ref{GP2}) has been obtained after a rescaling of temporal and spatial variables
as well as the magnitude of the order parameter and temperature by coefficients that are proportional to powers of $\tau _{q}$. Its kinetic and potential energy do not depend anymore on $\tau _{q}$. The rescaling factors $\alpha$ and $\beta$ directly relate to the ``freeze-out'' time and ``freeze-out'' length of the traditional KZ theory. The freeze-out time is $\hat{t}$ and freeze-out length is $\hat{\xi}$
\begin{align}
 \hat{t}\propto\alpha^{-1} & = \tau_q^{n/(n+1)} \label{eq:KZt}\\
  \hat{\xi}\propto\beta^{-1}& = \tau_q^{n/(2(n+1))}. \label{eq:KZxi}
\end{align}
For linear quenches $\hat{t}=\tau_q^{1/2}$ and $\hat{\xi}=\tau_q^{1/4}$, in agreement with KZ theory prediction with critical exponents $\nu=1/2$ and $z=2$ \cite{PhysRevLett.78.2519}.  The rescaling equations (\ref{rescaling1})-(\ref{rescaling2}) brings out the universality of the order parameter as in the rescaled coordinates the dynamics is identical for different quench rates.
This is the main idea of the present Letter as now the
question of scaling in the quench time $\tau _{q}$ has been transferred from the analysis of the dynamics of
a complicated equation to the scaling behavior of the parameter transformation that yields a quench
rate independent equation of motion. The Kibble-Zurek theory of topological defect formation then
follows directly from this result if we assume that topological defects are formed in the system
upon traversal of the critical point.

The linear transformations that resulted in a $\tau_q$-independent dynamical equation (\ref{GP2}) involved a rescaling of temperature. This suggests that, formally, the scaling relations (\ref{eq:KZt})-(\ref{eq:KZxi}) hold in experiments, where temperature is varied to keep $T\beta^{d-4}$ constant.
Typically, however, the KZ experiments are performed in systems that interact with a heat reservoir at a constant temperature, which is independent of quench rate. Thus to evaluate the universality of the above approach we consider the effect of the rescaling of the stochastic term in equation (\ref{GP2}). The rescaling $T\rightarrow \tilde{T}= T\beta^{d-4}=T\tau_q^{-\frac{n(d-4)}{2(n+1)}}$ affects the initial conditions and also changes the amount of thermal energy introduced into the system during the quench. For $d=4$ the temperature dependence on quench rate disappears and equation (\ref{GP2}) is  universal. Now let us consider how the temperature affects quench dynamics in dimensions $d\neq4$.  If the nonlinear term in the equations of motion can be neglected then the initial conditions are independent of temperature and hence quench rate. This is because, in this case, the order parameter can simply be rescaled to remove the temperature dependence. If the initial conditions are independent of quench rate and the thermal energy introduced by the stochastic term is small then the dynamics is universal. Let $\eta_0$ denote the time when the quench starts. Comparison of terms in equation (\ref{GP2}) suggests that the nonlinear term can be neglected when

\begin{equation}
|\eta_0^n\varphi|\gg|g\varphi^3|. \label{eq:eta0_1}
\end{equation}
In thermal equilibrium, the magnitude of the order parameter field scales with $\sqrt{\tilde{T}}$ i.e. $|\varphi|\sim \sqrt{\tilde{T}}$ and substituting this expression in (\ref{eq:eta0_1}) gives

\begin{align}
|\eta_0| & \gg  A,  \nonumber \\
A &\equiv \left| g T \tau_q^{-\frac{n(d-4)}{2(n+1)}}\right|^{\frac{1}{n}}. \label{eq:Acond}
\end{align}
Thus as long as the condition (\ref{eq:Acond}) is valid the KZ scaling given by eqs. (\ref{eq:KZt})-(\ref{eq:KZxi}) holds. Let us consider in which cases the condition (\ref{eq:Acond}) is satisfied. Since the KZ scenario is concerned with scaling at slow quenches we examine the slow quench limit. For $d>4$, reducing the quench rate $1/\tau_q$ decreases $A$. Thus when $d\geq 4$ the condition (\ref{eq:Acond}) can be always satisfied for slow quenches and the dynamics is universal. Dimensions $d < 4$ require a more careful treatment because condition (\ref{eq:Acond}) will be eventually violated at very slow quenches at which point the fluctuations will start to influence the KZ scaling to destroy the behaviour that would be expected from universal equations. By analogy with the equilibrium thermodynamics we may regard dimension $d = 4$ as an upper critical dimension. Nevertheless, for any finite temperature $T$ there will be a range of quench rates $1/\tau_Q$ such that eq. (\ref{eq:Acond}) is satisfied and in this range the dynamics will reproduce the behaviour of a universal equation. One should note that real systems are finite and for slow quenches the finite size effects will also become significant and can easily dominate the observed scaling \cite{PhysRevLett.109.015701}.

In practice, KZ experiments measure the density of stable defects $\rho(\textbf{r})$ at the end of the quench protocol. In $d$-dimensional space, $\rho({\mathbf{r }})$ and $\rho({\boldsymbol{\xi }})$
in the original and the rescaled coordinates, respectively, satisfy
\begin{equation}
    \int \rho({\mathbf{r }}) d^d{\mathbf{r }} = \int \rho({\boldsymbol{\xi }}) d^d{\boldsymbol{\xi }}. \label{integral}
\end{equation}
Making use of equations (\ref{rescaling1}) and (\ref{rescaling2}) in (\ref{integral}) we
find that the defect density in real space scales as
\begin{equation}
    \rho_{def} \sim \beta^{d}\sim \tau _{q}^{-\frac{n}{2(n+1)}d}.  \label{scaling_law}
\end{equation}
Alternatively, this result can be formulated for characteristic domain size, which scales linearly
with $\left\vert \mathbf{r}\right\vert $. Thus in real space the characteristic domain size scales
as $\sim \beta ^{-1}$. This is in agreement with predictions of the Kibble-Zurek
theory, e.g. for $n=1$, $d=1$, we find $N_{def}\sim \tau _{q}^{-1/4}$ which agrees with the results presented
in \cite{PhysRevLett.78.2519,PhysRevLett.80.5477,PhysRevLett.104.160404}.

It is not possible to have arbitrarily many defects in the system
and therefore at high quench rates (high defect densities) one expects to observe a plateau in the defect
scaling. This deviation from the simple power-law scaling can be anticipated by noting
the scaling of the characteristic defect size and comparing it to the scaling of separation between defects.
The characteristic size of the defects in rescaled coordinates is given by $\tilde{h}\sim\sqrt{2/g|\varphi|^2}$, as according to equation (\ref{GP2}), a shorter defect has a kinetic energy that exceeds the potential barrier. By making use
of the relations (\ref{rescaling1}) one can see that in real space the typical defect
size $h$ has no dependence on $\tau_{q}$. On the other hand the typical
distance between defects, i.e. the domain size, scales as $D\sim\beta^{-1}$.
The simple power law scaling breaks down when the ratio $h/D \sim \beta$ becomes of order of one, which happens for sufficiently small $\tau_q$.
We emphasize that this physical argumentation is not employed in our derivation of the power law-scaling, but is simply used to anticipate its breakdown
at fast quench rates.

{\em The spatially inhomogeneous case --} Now we will apply the same strategy to analyze the system in the
presence of space-dependent external potential $V\left( \mathbf{r},L\right)$. In this case
the dynamics of the system is governed by the equation%
\begin{align}
    \frac{\partial \phi}{\partial t} &= -\Gamma \left[ \nabla ^{2}%
    \phi-\left\vert \dfrac{t}{\tau _{q}}\right\vert ^{n}\mathrm{sign}%
    \left( t\right) \phi+g\left\vert \phi\right\vert ^{2}\phi%
    -V\left( \mathbf{r},L\right) \phi\right]\nonumber\\
    & +\theta\left( \mathbf{r} ,t\right) \sqrt{T}.  \label{GP_trap}
\end{align}
which contains an additional spatially dependent term. Therefore the dependence on $\tau _{q}$
can in general be eliminated only if the trapping potential $V\left( \mathbf{r},L\right)$ is
also rescaled. To be specific let us consider here two such potentials:

(i) power law potential $V\left(r,L\right) =\left\vert \frac{r}{L}\right\vert ^{m}$, with $r = \left| \mathbf{r}  \right|$. The dependence
on $\tau _{q}$ in eq. (\ref{GP_trap}) is eliminated if in addition to rescalings (\ref{rescaling1}) and
(\ref{rescaling2}) one assumes that the characteristic system size is rescaled according to $\lambda = \beta \alpha ^{1/m} L$. Then the equation of motion (\ref{GP_trap}) is reduced to the
following $\tau _{q}$-independent universal equation%
\begin{align}
    \frac{\partial \varphi }{\partial \eta } &= -\Gamma \left[ \nabla
    _{\xi }^{2}\varphi-\mathrm{sign}\left( \eta \right) \left\vert \eta
    \right\vert ^{n}\varphi+\left\vert \frac{\xi }{\lambda }\right\vert
    ^{m}\varphi+g\left\vert \varphi\right\vert ^{2}\varphi%
    \right]\nonumber\\
    & +\theta\left( \boldsymbol{\xi },\eta \right) \sqrt{\widetilde{T}}.
    \label{GP_trap2}
\end{align}

(ii) For an inverted Gaussian potential $V\left( x,L\right) =\exp \left( -\frac{x^{2}}{%
L^{2}}\right) $ the equation of motion (\ref{GP_trap}) can be brought to universal
form if the following nonlinear rescaling is applied $L^{2}=-\frac{\xi ^{2}}{\beta ^{2}
\ln \left( \lambda \alpha \right) }$. This allows us to write the equation of motion
in the universal form
\begin{align}
    \frac{\partial \varphi_{0}}{\partial \eta } &= -\Gamma \left[
    \nabla _{\xi }^{2}\varphi-\mathrm{sign}\left( \eta \right) \left\vert
    \eta \right\vert ^{n}\varphi-\lambda \varphi+g\left\vert
    \varphi\right\vert ^{2}\varphi\right]\nonumber\\
    & +\theta\left( \boldsymbol{\xi},\eta \right) \sqrt{\widetilde{T}}
\end{align}

These examples illustrate that in inhomogeneous settings the number of topological defects depends not only on the quench rate but also on the characteristic system size $L$.
 i.e. in the rescaled frame of reference we can write $\rho({\boldsymbol{\xi }}, \lambda)$. Thus in original space for the density of defects we obtain the following relation $\rho({\mathbf{r}}, \lambda\left(L,\tau_q \right))\sim \beta^{d}\sim \tau _{q}^{-\frac{n}{2(n+1)}d}$ provided that $\lambda$ is kept constant.
This result differs from the standard inhomogeneous KZ theory based on intuitive arguments. In particular, it was predicted that the influence of external potential may change the power of dependence of $\rho$ on $\tau_q$ \cite{PhysRevLett.102.105702,PhysRevLett.105.075701}. However, as it was shown above the density of defects will follow a precise powerlaw scaling in $\tau_q$ \emph{only} if the characteristic system size $L$ is varied in such a way that $\lambda(L,\tau_q)$ does not depend on $\tau_q$. The required transformation $L(\tau_q)$ can be determined from the equations of motion, as it has been demonstrated in the two examples.

{\em Beyond Ginzburg-Landau --}
Our method is based on the derivation of universal equations by means of a rescaling of
spatial and temporal coordinates. It can also be applied to the systems whose dynamics is
described by different equations of motion describing phase transition of systems in other
dynamic universality classes \cite{RevModPhys.49.435}.
Naturally, in this case the specific form of rescaling coefficients in (\ref{rescaling1})
and (\ref{rescaling2}) is modified. However, if the equation
of motion is known, the method enables the simple derivation of number of defects scaling as a
function of $\tau _{q}$. In order to illustrate this let us analyze the dynamics of the
system in the so-called underdamped regime \cite{PhysRevD.58.085021}. The dynamics of the
field in this case is governed by equation
\begin{equation}
    \frac{\partial ^{2}\phi}{\partial t^{2}}= \nabla ^{2}%
    \phi-\left\vert \dfrac{t}{\tau _{q}}\right\vert ^{n}\mathrm{sign}%
    \left( t\right) \phi+g\left\vert \phi\right\vert ^{2}\phi.
     \label{Ion}
\end{equation}%
Equation (\ref{Ion}) differs from (\ref{GP}) in that the first time derivative is
replaced by a second time derivative.
Now the quench time in (\ref{Ion}) is eliminated if the
following rescalings are applied%
\begin{align}
    \eta  &=\alpha t,\text{ }\boldsymbol{\xi }=\beta \mathbf{r},
    \label{Ion_scaling} \\
    \alpha  &=\tau _{q}^{-\frac{n}{n+2}}, \text{ }\beta=\alpha. 
\end{align}
Thus in the linear quench regime $(n=1)$ and for $d=1$, the defect density scales
according to $\rho(\mathbf{r}) \sim \beta \sim \tau _{q}^{-\frac{1}{3}}$. This result is
in perfect agreement with the analytical and numerical results presented in \cite{PhysRevD.58.085021,PhysRevLett.80.5477,Nigmatullin2016}.

It should be emphasized that the transformation approach that we have described in this Letter
is not restricted to equations of motion of order parameters but does apply equally to the
full microscopic equations of motion that are underlying the system. For instance by making use of the methods presented here one can rewrite the equation of motion of a quantum Ising chain in a universal form by assuming $\alpha=\tau _{q}^{-1/2}$, $\beta=\alpha$ \cite{PhysRevLett.109.015701}.

{\em Conclusions--} We have presented an analytical method for examining the
dynamics of second-order phase transitions near the critical point that replaces physical arguments
by mathematical reasoning based on transformations of the equations of motion of a system under
consideration. The power of the method is demonstrated by considering two specific cases of the time
dependent Landau-Ginzburg equation in the so-called overdamped and underdamped regimes. We have shown
that by making use of linear transformations the equation of motion can be represented in a universal
form with no dependence on quench rate. This has been used to derive the spatial scaling of the defect
density with the quench rate of the transition. 
We have also applied our method to analyze the dynamics of inhomogeneous systems and have shown
that in this case the number of defects also scales with the characteristic system size. The approach presented here goes well beyond these problems and can be applied to the microscopic equations of motion underlying arbitrary phase transitions.

{\em Acknowledgments --} We acknowledge discussions with A. del Campo, A. Retzker, M. Bruderer. This
work was supported by an Alexander von Humboldt Professorship, the EPSRC Doctoral Training
Center for Controlled Quantum Dynamics and EPSRC National Quantum Technology Hub
in Networked Quantum Information Processing, the EU Integrating Project SIQS and the EU STREP
projects PICC and EQUAM.

\end{document}